\documentclass[%
 reprint,
 amsmath,amssymb,
 aps,
 prd,
]{revtex4-2}

\usepackage{graphicx}
\usepackage{dcolumn}
\usepackage{bm}

\usepackage{siunitx}

\newcommand{\grammage}{\unit{g.cm^{-2}}}

\begin{document}


\title{Anatomy of an extensive air shower:\\building an optimal radio emission calculation}

\author{Juan Ammerman-Yebra}
\author{Harm Schoorlemmer}
 \affiliation{IMAPP, Radboud University Nijmegen, Nijmegen, The Netherlands}
\affiliation{Nationaal Instituut voor Kernfysica en Hoge Energie Fysica (NIKHEF), Science Park, Amsterdam, The Netherlands}




\newcommand{\juan}{\textcolor{red}}

\date{\today}

\begin{abstract}
Radio emission from particle air showers is one of the most promising detection techniques to achieve an experiment with a large exposure in ultra-high-energy astroparticle physics. In this regime, above $10^{17}$ eV, the detailed calculation of the radio emission from Monte Carlo particle showers is sufficiently CPU intensive that statistical studies and optimization problems remain largely out of reach. In this article we examine the characteristics of the longitudinal and lateral development of extensive air showers, and exploit them to obtain a fast evaluation of the associated radio emission. We introduce the Coherent Radio Emission from Positrons and Electrons in Showers (CREPES) algorithm, in which the particle distributions are precomputed once and the emission is subsequently obtained from them, yielding a gain in speed of more than three orders of magnitude while preserving the accuracy of the microscopic calculation.
\end{abstract}

\maketitle


\section{\label{intro} Introduction}

The detection of radio-wave emission from particle showers has become one of the most promising techniques in ultra-high-energy (UHE) astroparticle physics, allowing both the detection of the shower and the reconstruction of the properties of the primary particle \cite{Schroder:2016hrv}. This is due to the long attenuation length of radio waves in air, or ice, together with the comparatively low cost of MHz-range antennas, which together make radio detection the most viable option currently available with which to instrument the large collecting areas needed to measure the low fluxes of UHE neutrinos, photons, and cosmic rays.

As the primary particle interacts in the dense medium, it produces secondary particles that continue to interact and multiply, building up the particle cascade. Since the medium itself is populated with atomic electrons, a fraction of these are swept up and accelerated into the shower, so that the cascade develops a net negative charge excess. These particles travel at velocities close to the speed of light and this charge excess behaves effectively as a negative charge moving coherently through the medium. It is this motion that gives rise to the radiation known as the Askaryan mechanism \cite{Askaryan:1961pfb}, characterized by an electric field radially polarized around the shower axis and beamed sharply along the Cherenkov cone.

If the dense medium has a low density (or a sufficiently strong magnetic field), like the atmosphere, the presence of a magnetic field will generate a drift in the charged particles. 
Electrons and positrons, being of opposite charge, are deflected in opposite transverse directions, so that an effective transverse current builds up across the shower front and propagates at nearly the speed of light \cite{Scholten:2007ky}. The resulting radiation is polarized along $\mathbf{v}\times\mathbf{B}$, where $\mathbf{v}$ denotes the shower direction and $\mathbf{B}$ the local geomagnetic field. For showers developing in the Earth's atmosphere this geomagnetic emission is typically the dominant contribution as both electrons and positrons participate in it constructively, in contrast with the Askaryan mechanism, to which only the net charge excess contributes.

The electric field at a given observer position can in principle be obtained by summing the contributions of every individual particle track generated by a Monte Carlo shower code such as Aires \cite{sciutto1999aires} or \textsc{CORSIKA}7 \cite{Heck:1998vt}. This is the strategy followed by ZHS\cite{zas1992electromagnetic}, by the Endpoints formalism \cite{James:2010vm}, and by Eisvogel\cite{Windischhofer:2023ahw}. Since both the field amplitude and the propagation time have to be evaluated separately for each track, the computational cost of this approach grows with the number of particles in the shower and hence, with the primary energy. Thinning \cite{Hillas:1997tf} alleviates this cost by statistically sampling the secondary particles, discarding a fraction of the particles while increasing the statistical weight of those retained. This comes at the price of introducing coherence artifacts that limit the accuracy of the resulting pulse. Eisvogel instead sidesteps part of the per-track propagation calculation by precomputing the Green's function of the medium. This precomputation is itself expensive but needs to be performed only once for a given medium, after which the contribution of each individual track can be evaluated more efficiently.

A number of semi-analytical methods have since been developed that exploit the structure of the particle distributions in order to accelerate the radio emission calculation. For showers in ice, the ARZ method \cite{Alvarez-Muniz:2020ary} achieves this by assuming a homogeneous medium and a far-field observer at the Cherenkov angle. For this observer, the entire longitudinal development of the shower is seen at once, allowing the radial charge distribution to be inferred from the simulated radio pulse and subsequently reused to compute the emission at any other viewing angle. For air showers, EVA \cite{Werner:2012cr} and MGMR3D \cite{Scholten:2017tcr} instead parametrize the particle distributions directly from Monte Carlo simulations and compute the electric field from the effective transverse current by integrating over shower bins rather than over individual tracks. This last approach, while producing very promising results, still relies on some simplifications of the shower structure. The main one being the assumption that particles are symmetrically located around the shower axis, which make it a useful tool for many applications \cite{Trinh:2020gce, Mitra:2023zjf, Scholten:2024upn} but not yet accurate enough to substitute full, detailed simulations.

An intermediate approach between the two schemes discussed above was proposed in \cite{Cummings:2023tuh}. The information contained in a simulated particle shower is compressed onto a four-dimensional grid, so that all the tracks going through a given cell are represented by a single average track. This allows the construction of a library of compressed showers from which the radio emission can be evaluated rapidly, without the need to rerun the particle shower simulation. The approach is, however, subject to a number of caveats: any change in the hadronic interaction model, in the frequency range under consideration (provided that the grid was not chosen sufficiently fine at first) or in the experimental site would require the regeneration of the library, and the emission for a given shower geometry can only be obtained once the corresponding simulation has been run.

A different strategy altogether, followed by Radio-Morphing \cite{zilles2020radio, chiche2022radio} and  \cite{Corstanje:2023vqp}, is to interpolate between a precomputed library of simulated radio pulses in order to predict the emission at any observer positions. The main limitation of this approach is the upfront cost of generating the library. Such a library requires simulating a large number of observers and, if there was any change to an experimental site or hadronic model, it would require the library to be fully regenerated from scratch.

While the ARZ method has been able to solve the computational problem for in-ice particle showers and enabled the optimization based reconstructions and detector layouts (see, for example, \cite{Heyer:2024nkq, Heyer:2025cpw, Ravn:2025jzs}), the current approaches for extensive air showers still fall short of this goal, either because of insufficient accuracy or a lack of flexibility in the simulated particle showers. In this work we propose a new approach to calculate radio emission from extensive air showers, although the framework would also apply to in-ice showers, that will hopefully solve the above-mentioned problems. In Section~\ref{sec:radio} we revisit the conditions under which the radio emission from a particle shower is coherent, and identify the physical quantities on which this coherence depends. In Section~\ref{sec:shower_anatomy} we examine the particle distributions that develop within an air shower and use their structure to motivate the construction of a fast radio emission model. In Section~\ref{sec:crepes_recipe} we derive this model, to which we refer as CREPES (Coherent Radio Emission from Positrons and Electrons in Showers), and describe the algorithm by which it is evaluated. Lastly, in Section~\ref{sec:crepes_tasting} we test the accuracy of CREPES against detailed ZHAireS simulations \cite{AlvarezMuniz2012}, first for a vertical shower and then for an interpolated inclined shower.

\section{\label{sec:radio} Radio emission from particle showers}

To understand the radiation from a particle shower we start from the Maxwell equations in the transverse gauge, written for a linear, isotropic, and non-dispersive medium \cite{Jackson:1998nia}:
\begin{equation}
    \nabla^2\phi = -\frac{\rho}{\epsilon}
\end{equation}
\begin{equation}
    \nabla^2\mathbf{A}-\mu\epsilon\frac{\partial^2\mathbf{A}}{\partial t^2} = -\mu\mathbf{J}_\perp
\label{eq:maxwell_vector_potential}
\end{equation}
where $\phi$ and $\mathbf{A}$ are the scalar and vector potentials, $\rho$ is the charge density of the source, and $\mathbf{J}_\perp$ is the transverse current, the divergence-less part of the current density, which for an observer at large distances from the source reduces to the component of $\mathbf{J}$ perpendicular to the line of sight
$\hat{\mathbf{u}}$, i.e.\ $\mathbf{J}_\perp = -\,\hat{\mathbf{u}}\times(\hat{\mathbf{u}}\times\mathbf{J})$. For an observer in the far-field regime it can be shown that only the vector potential contributes to the radiated field \cite{Alvarez-Muniz:2010wjm}. Solving eq.~\eqref{eq:maxwell_vector_potential} with Green's function yields:
\begin{equation}
\begin{split}
    \mathbf{A}(\mathbf{x},t) =& \frac{\mu_0}{4\pi}\iint \frac{\mathbf{J}_\perp(\mathbf{x}', t')}{|\mathbf{x}-\mathbf{x}'|}\,\\
    &\times \delta\!\left(t' - t + \frac{n_\text{eff}}{c}|\mathbf{x}-\mathbf{x}'|\right)\,d^3\mathbf{x}'\,dt'\\
\label{eq:vector_potential_sol}
\end{split}
\end{equation}
where $\mathbf{x}'$ and $t'$ are the position and time of the source point, $\mathbf{x}$ and $t$ those of the observer, and $n_\mathrm{eff}$ is the refractive index averaged along the optical path connecting the two. The electric field is then obtained from the time derivative, $\mathbf{E}=-\partial\mathbf{A}/\partial t$.

Two terms therefore determine eq.~\eqref{eq:vector_potential_sol}: the transverse current $\mathbf{J}_\perp$ and the delta function. The delta function assigns to every source point the observer time at which its radiation arrives, so that emission generated at different places and at different times throughout the shower can be mapped onto the same interval of the observer time. The transverse current, in turn, sets the weight of each of these contributions and, being the only vector quantity appearing in the integrand, also fixes the polarization of the resulting pulse. The value of $\mathbf{A}$ at a given observer position and time thus depends both on how many particles are present at each source point and on whether that point shares its arrival time with other regions of the shower.

\subsection{The transverse current}
The transverse current is built from the positions and velocity vectors of the individual charges that make up the shower. Electrons and positrons traveling side by side in the same direction carry currents of opposite sign that cancel, so that a shower composed of $e^+e^-$ pairs would produce no net transverse current and would not radiate. Two physical effects break this symmetry.

The first is the charge excess. Electrons of the medium are swept into the cascade as it develops, while positrons are removed by annihilation, and the shower ends up carrying a net negative charge that moves with the shower front. This is the origin of the Askaryan emission \cite{Askaryan:1961pfb}.

The second is the deflection in the Earth's magnetic field. The Lorentz force pushes electrons towards one side of the shower and positrons towards the other, so that charges of opposite sign drift in opposite directions and their currents add instead of canceling. The result is a transverse current that propagates with the shower front nearly at the speed of light \cite{Werner:2007kh}.

In both cases the transverse current is governed by where the particles are as the shower develops, by the velocity components they acquire through scattering, and by how far they drift between interactions. Following the position and the direction of motion of the charges is therefore a prerequisite for evaluating the vector potential.

\subsection{Coherence conditions}

In a shower containing billions of particles that move in a similar direction, many
source points satisfy the delta function of eq.~\eqref{eq:vector_potential_sol} for
the same observer and the same observer time. When the fields radiated by these
points add in phase, we speak of coherence. The width of the time interval within
which contributions are treated as simultaneous fixes the frequency up to which the
emission remains fully coherent, and conversely, a model that aims to reproduce the
field up to a frequency $\nu$ must describe the timing of the particles down to
$t\approx 1/2\pi\nu$.

\begin{figure}[!ht]
\includegraphics[width=\linewidth]{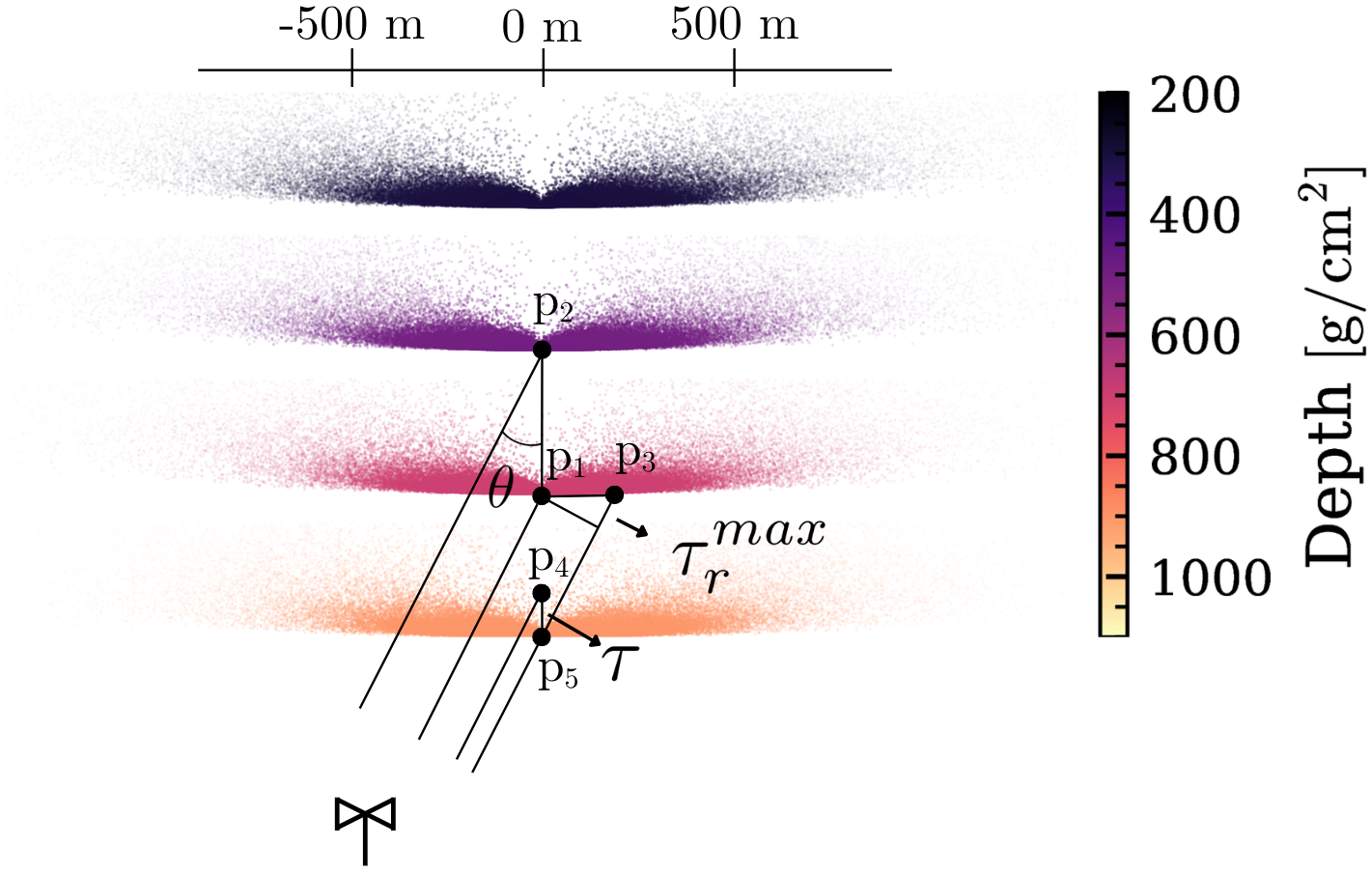}
\caption{\label{fig:sfd_sketch} Shower front snapshots at four development stages
(300, 500, 700 and 900~\grammage) for a 1~EeV proton vertical shower. Particles are
recorded when they cross a plane at a given depth, and the vertical axis is their
shower front delay converted to distance by multiplying by the speed of light.}
\end{figure}

Figure~\ref{fig:sfd_sketch} shows snapshots of a $1$~EeV vertical shower at four
stages of its development, obtained by recording every particle that crosses a plane
perpendicular to the shower axis together with its position, direction of motion and
crossing time. The vertical axis of each snapshot is the delay of a particle with
respect to the first one to cross the plane, converted into a distance by multiplying
by the speed of light. The marked source points $p_1$ to $p_5$ illustrate the
different ways in which two emission regions can be displaced with respect to each
other: along the shower axis, within the shower front, and in radial distance from
the axis.

Each of these displacements translates into a delay at the observer, and three
conditions must be satisfied for two regions of the shower to contribute coherently
to the field observed at a given position \cite{Ammerman-Yebra:2023rhr}:
\begin{itemize}
    \item[-] Cherenkov condition: $z(1-n\beta\cos\theta)/c<\frac{1}{2\pi\nu}$
    \item[-] Shower front delay condition: $\tau<\frac{1}{2\pi\nu}$
    \item[-] Radial position condition: $\tau_r=|nr\sin\theta\cos\phi|/c<\frac{1}{2\pi\nu}$
\end{itemize}
where $\theta$ is the angle to the observer, as defined in
Fig.~\ref{fig:sfd_sketch}, and $r$ and $\phi$ are, respectively, the radial distance
from the shower axis and the azimuthal angle around it at which the source point is
located.

The Cherenkov condition compares emission from points that occupy the same position
in the shower front but belong to different stages of the development, such as $p_1$
and $p_2$. For two such points separated by a distance $z$ along the axis, the
emission reaches the observer with a time difference $z(1-n\beta\cos\theta)/c$. This
difference vanishes at the Cherenkov angle, $\theta_\mathrm{C}=\arccos(1/\beta n)$,
where the whole longitudinal development contributes in phase. This last statement is
only partially true, but still useful for the general understanding, as an extensive
air shower develops in an inhomogeneous atmosphere where each stage of the shower has
a different Cherenkov angle and not all particles travel at nearly the speed of light.

The shower front delay condition applies to particles that belong to the same
development stage but cross a given plane at different times, such as $p_4$ and
$p_5$. The particle at $p_5$ crosses at $t=0$ and its emission needs a time
$n_\mathrm{avg}R/c$ to reach the observer. The particle at $p_4$ reaches the same
plane $\tau$ later, and its emission travels along essentially the same path. The
observer receives the two contributions separated by $\tau$, which limits coherence
to frequencies below $1/2\pi\tau$. Reproducing the spread of the shower front is thus
essential for the coherence of the final pulse.

The radial condition tells which particles at different radial positions within the
same shower front contribute coherently. In the far field the extra path introduced
by a radial offset $r$ is $nr\sin\theta\cos\phi/c$, so particles lying along the line
perpendicular to the observer direction, $\phi=90^\circ$ or $\phi=270^\circ$, radiate
with no relative delay, while particles at any other azimuth arrive earlier or later
than one on the shower axis. In Fig.~\ref{fig:sfd_sketch} the emission of $p_3$
reaches the observer $nr\sin\theta/c$ after that of $p_1$, which is the maximum
radial delay $\tau_r^{max}$ for that offset.

\section{\label{sec:shower_anatomy}Particle distributions in  extensive air showers}

It follows from the discussion above that a calculation of the radio emission from a shower requires detailed knowledge of how its particles are distributed in space and time. To illustrate this we consider the case of a vertical shower developing along $z$ axis, of primary energy 1 EeV, propagating in the Earth's atmosphere with a horizontal magnetic field oriented along the $x$ axis. To obtain smoother distributions, all the results shown are averages over 20 showers simulated with Aires.

\begin{figure}[!ht]
\includegraphics{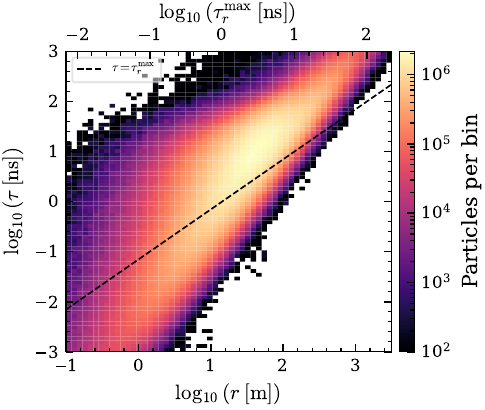}
\caption{\label{fig:max_shower_particle_distribution} Average distribution of electrons and positrons at 700 g/cm2 for a 1 EeV proton shower as a function of radial position and shower front delay. The upper $x$-axis corresponds to the maximum radial delay for a given radial position for an observer placed at the Cherenkov angle. The dotted black line delimits the points in which particles have equal shower front delay and maximum radial delay.
}  
\end{figure}

We have seen in Section~\ref{sec:radio} that the radial position and the shower front delay of a particle together determine whether it contributes coherently to the field seen by a given observer. Fig.~\ref{fig:max_shower_particle_distribution} shows the particle distribution as a function of these two delays at a depth of 700 \grammage, close to shower maximum. The horizontal axis gives the radial position of the particles, integrated over all azimuthal angles, with the corresponding maximum lateral delay, $\tau_r^\mathrm{max}=nr\sin\theta/c$, shown on the upper axis. The black line separates the region in which the shower front delay exceeds the maximum radial delay (above the line) from that in which the opposite holds (below the line).

\begin{figure}[!ht]
\includegraphics[width=0.48\textwidth]{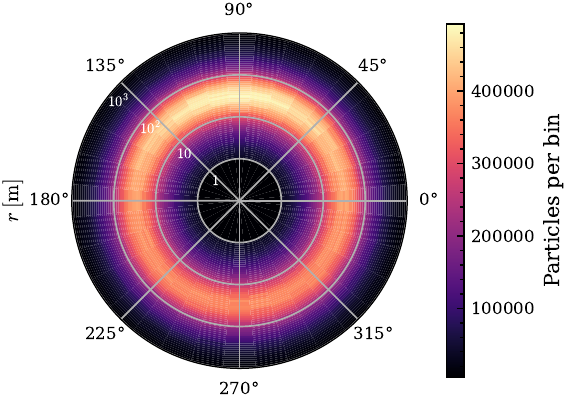}
\caption{\label{fig:azimuthal_dist_at_xmax_vertical} Average distribution of electrons and positrons at 700 g/cm2 for a 1 EeV proton shower as a function of radial position and azimuth angle. The inner bin ($r<1$ m) contains all particles below that radius.
}
\end{figure}

Fig.~\ref{fig:max_shower_particle_distribution} shows that the two types of delay interplay with one another rather than acting independently. For showers in air, the shower front delay is the dominant contribution to the coherence of the radio pulse, since the bulk of the particle distribution lies above the black line. A non-negligible fraction of particles still lies below it, which indicates that beyond 100 MHz the radial position begins to play a role as well (for a more detailed discussion see \cite{Ammerman-Yebra:2023rhr} and \cite{AmmermanYebra:2025mir}).

In Fig.~\ref{fig:azimuthal_dist_at_xmax_vertical} we show the angular distribution of the particles at the same depth of 700 \grammage, having now compressed the shower front delay distribution in order to expose any azimuthal asymmetry. For this geometry a clear excess of particles is visible at $\phi=90^\circ$, which corresponds to the electrons drifting under the influence of a magnetic field pointing along the $x$ axis while the shower itself propagates toward negative $z$.

\begin{figure}[!ht]
\includegraphics{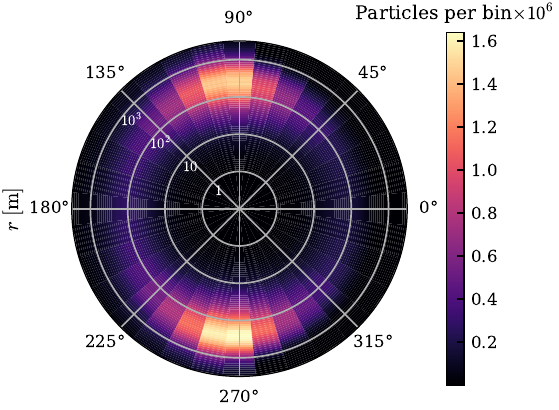}
\caption{\label{fig:azimuthal_dist_at_xmax_80} Same as Fig.~\ref{fig:azimuthal_dist_at_xmax_vertical} for an 80 degree 1 EeV proton shower developing in Earths atmosphere with a vertical magnetic field.
}
\end{figure}

The vertical shower represents the case in which particles are deflected least, since it rapidly penetrates the atmosphere and develops within its densest layers. Fig.~\ref{fig:azimuthal_dist_at_xmax_80} shows the azimuthal particle distribution at the same depth of 700\grammage{} for a 1 EeV shower of 80$^\circ$ zenith angle, propagating through the atmosphere under a vertical magnetic field. Under this configuration the change in field direction shifts the maxima of the particle distribution, with electrons now drifting toward $\phi=270^\circ$ and a second peak appearing on the opposite side of the distribution, produced by the positrons, since the larger deflection distances available to particles at this zenith angle allow them to separate more clearly. 
From these results we conclude that the assumption of a radially symmetric particle distribution does not hold for extensive air showers.

\begin{figure}[!ht]
\includegraphics[width=\linewidth]{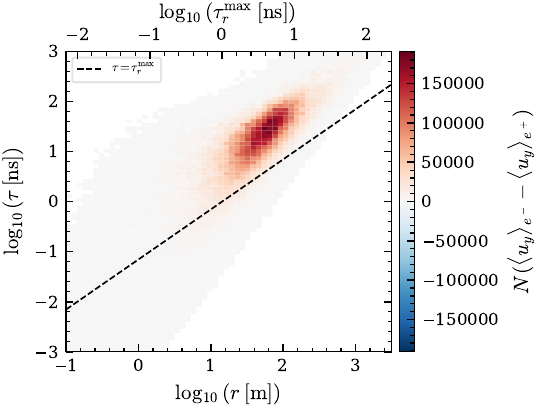}
\caption{Visualization of the geomagnetic current density terms as a function of radial position in the shower axis and shower front delay. Distribution taken at 700 g/cm2 for a 1 EeV vertical proton shower.}
\label{fig:uy_velocity_distribution}
\end{figure}

We state in Section~\ref{sec:radio} that the current density depends not only on where the particles are located but also on their direction of motion, since it is this direction that determines how the current builds up. The geomagnetic current arises from the drift of electrons and positrons in opposite transverse directions. For the vertical shower with a horizontal magnetic field, electrons drift toward positive $y$ while positrons drift toward negative $y$. Fig.~\ref{fig:uy_velocity_distribution} visualizes this current by weighting the particle distribution of Fig.~\ref{fig:max_shower_particle_distribution} with the average $y$ component of the velocity of electrons minus that of positrons. Most of the particles contributing to the transverse current are seen to have delays between 10 and 100 ns. The fact that this distribution lies above the black line indicates that, for the geomagnetic emission, the shower front delay dominates over the radial delay. This indicates that it is possible to model the geomagnetic emission to first order at low frequencies using a one-dimensional shower that accounts only for the spread of particles along the shower front \cite{ammerman2026ARENA}.

\begin{figure}[!ht]
\includegraphics[width=\linewidth]{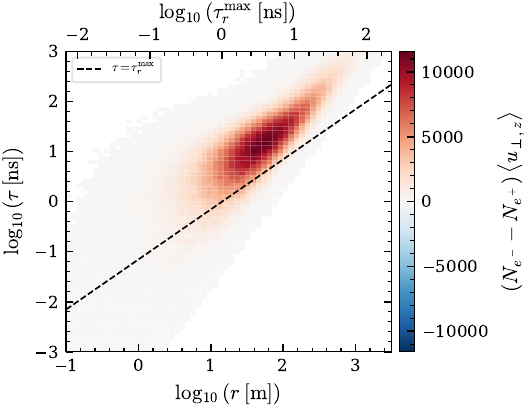}
\caption{Same as Fig.~\ref{fig:uy_velocity_distribution} for the negative charge excess terms.}
\label{fig:ch_excess_distribution}
\end{figure}

To visualize the Askaryan radiation, Fig.~\ref{fig:ch_excess_distribution} shows the particle distribution of Fig.~\ref{fig:max_shower_particle_distribution} weighted instead by the charge excess fraction and by the projection of the $z$ component of the velocity onto the direction transverse to the line of sight. The resulting distribution extends over a substantially larger range of shower front delays than the geomagnetic case, and, in particular, extends visibly below the black line. The radial delay is therefore important for the Askaryan mechanism, so that a purely one-dimensional model would not suffice to describe it. This behavior differs from that of the geomagnetic emission because particles need not be deflected in order to radiate via the Askaryan mechanism, and consequently begin contributing to the current already at smaller shower front delays.

\section{\label{sec:crepes_recipe}The CREPES recipe}
The results of the previous section show that a correct calculation of the radio pulse from a shower requires the particle distributions to be known in three-dimensional space and resolved in time to nanosecond accuracy. Traditional Monte Carlo codes achieve this by following every individual particle track through the shower, a procedure that is computationally very costly. Thinning, a statistical technique that assigns a weight to each particle so that not all of them need be followed individually, can alleviate this cost to a certain degree, before coherence artifacts begin to appear. As the primary energy increases, however, so does the number of particles in the shower, and even thinning is unable to bring the simulation time for events with more than ten antennas below a day.

To circumvent this limitation we follow an approach already used in \cite{Werner:2007kh, WERNER20125, Scholten:2017tcr}, in which the radio emission is obtained not from individual particle tracks but by integrating directly the current densities within the shower. This means that the contribution of all the particles contained in a given volume bin at a certain time is evaluated simultaneously rather than one track at a time.

For the derivation that follows we consider, without loss of generality, a vertical shower advancing along the $z$ axis, and we address first the simpler case in which the entire shower is treated as developing along the shower axis itself. A fully three-dimensional calculation additionally requires binning the particle distributions in the radial and azimuthal coordinates, but each such bin then reduces to the same one-dimensional problem of projecting all its particles onto a single axis.

We start from eq.~\eqref{eq:vector_potential_sol} in order to derive the expression used in our calculation. Eq.~\eqref{eq:vector_potential_sol} is written in terms of a global source time, but, as established in Section~\ref{sec:shower_anatomy}, it is far more natural to describe the shower in terms of the shower front delay. We then introduce the change of variable:
\begin{equation}
    \tau = t' - t_{\textrm{front}}(z) \quad \Rightarrow \quad \textrm{d}\tau = \textrm{d}t'
\end{equation}
where $t_\mathrm{front}(z) = (h_\mathrm{start}-h(z))/c$. Projecting all particles onto the central axis of the bin under consideration, the current density can then be written as:
\begin{equation}
\begin{split}
    \textbf{J}_\perp =& q \frac{N(z')\,w(z',\tau)\textbf{u}_\perp(z',\tau)}{u_z(z',\tau)}\delta(x')\delta(y')
\end{split}
\end{equation}
where $N(z')$ is the number of particles crossing a plane perpendicular to the shower axis at position $z'$, $w(z',\tau)$ is the probability of finding a particle at $z'$ with shower front delay $\tau$, normalized so that $\int w(z',\tau)\,d\tau=1$, $\mathbf{u}$ is the unit velocity vector of the particles, and $q$ is the charge of the species being evaluated. It is the projection onto the central axis of the bin that gives rise to the factor $u_z(z',\tau)$ in the denominator. Since the particles are sampled as they cross a plane perpendicular to the shower axis, a condition that guarantees $u_z(z',\tau)\neq0$ for any particle so sampled, this term will not lead to a divergence. When $u_z(z',\tau)$ happens to be very small, the particle in question crosses the plane with a correspondingly large shower front delay, so that its significance to the radio pulse becomes negligible.

Substituting this current into eq.~\eqref{eq:vector_potential_sol} gives:
\begin{equation}
\begin{split}
    \mathbf{A}(\mathbf{x},t) =& \frac{\mu_0q}{4\pi}\iint \frac{1}{R}\frac{N(z')\,w(z',\tau)\textbf{u}_\perp(z',\tau)}{u_z(z',\tau)}\,\\
    &\times \delta\!\left(\tau-\tau_{\textrm{obs}}\right)\,dz'\,d\tau\\
\end{split}
\end{equation}
and discretizing the integral we arrive at
\begin{equation}
\begin{split}
\mathbf{A}(t_n) =& P \sum_{i_z=0}^{N_z-1}
\frac{\Delta z}{\tau_{i_z}(t_n) \cdot R_{i_z}}\\
&\left[
  \frac{N_p(X_{i_z})\,w_p\,\hat{\mathbf{u}}^{(p)}_\perp}{|\hat{u}_z^{(p)}|}
  - \frac{N_e(X_{i_z})\,w_e\,\hat{\mathbf{u}}^{(e)}_\perp}{|\hat{u}_z^{(e)}|}
\right]
\end{split}
\label{eq:1d-frape}
\end{equation}
where $P = \frac{e\mu_0}{4\pi}\cdot\frac{10^9}{\ln 10}$, the factor of $\ln 10$ arising because the shower front delays are integrated on a logarithmic scale in order to better characterize the distribution. It is worth noting that only the propagation time to each point in the $x$–$y$ plane of the shower front needs to be evaluated. This follows from the definition of the shower front delay itself, namely the additional time a particle takes to cross a plane perpendicular to the shower axis, relative to a particle travelling at the speed of light along that axis. Once the arrival time from a given point on the front is known, a particle with a shower front delay of $\tau$ ns simply contributes $\tau$ ns later than one with no delay at all. This considerably simplifies the calculation, since the propagation time needs to be computed only once per point on the axis rather than once per bin. The subscripts $e$ and $p$ denote electrons and positrons, respectively, whose contributions are kept separate in order to simplify the sampling of their distributions. It may also be noted that the particle velocity does not appear explicitly in eq.~\eqref{eq:1d-frape}. It is  accounted for implicitly through the evolution of the shape of the shower front delay distribution itself, which can be understood as encoding how fast or slow the particles are moving relative to a particle traveling at exactly the speed of light.

We argued in Section~\ref{sec:shower_anatomy} that a full three-dimensional description of the particle distribution is required in order to reproduce the radio pulse accurately. To incorporate this, we now apply the formalism derived above independently to each radial and azimuthal bin, so that eq.~\eqref{eq:1d-frape} generalizes to:
\begin{equation}
\begin{split}
\mathbf{A}(t_n) =& P \sum_{i_\phi=0}^{N_\phi-1}\;\sum_{i_r=0}^{N_r-1}\;\sum_{i_z=0}^{N_z-1}
\frac{\Delta z}{\tau_{i_\phi,i_r,i_z}(t_n) \cdot R_{i_\phi,i_r,i_z}}\\
&\left[
  \frac{N_p(X_{i_z})\,w_p\,\hat{\mathbf{u}}^{(p)}_\perp}{|\hat{u}_z^{(p)}|}
  - \frac{N_e(X_{i_z})\,w_e\,\hat{\mathbf{u}}^{(e)}_\perp}{|\hat{u}_z^{(e)}|}
\right]
\end{split}
\label{eq:crepes-algorithm}
\end{equation}

From now on we refer to this calculation as CREPES (Coherent Radio Emission from Positrons and Electrons in Showers). The workflow by which CREPES calculates the radio emission of a shower is done in four steps:
\begin{itemize}
    \item [1.] A set of showers with longitudinal profiles similar to that of the shower under study are simulated and selected. This step is in principle generalizable and will not be needed for every individual shower (see Section~\ref{sec:crepes_tasting80}).
    \item [2.] At a series of depth levels throughout each of these showers, the properties of every electron and positron crossing the plane perpendicular to the shower axis are recorded, namely the shower front delay, equivalently, the time at which the particle crosses the plane, together with its direction of motion and its position within that plane.
    \item [3.] These particle files are combined into the average look-up tables.
    \item [4.] Given a longitudinal profile for the shower under study, the emission is evaluated by interpolating the look-up tables and eq.~\eqref{eq:crepes-algorithm}.
\end{itemize}
Once assembled, this library of look-up tables allows the radio emission of a shower to be calculated without ever explicitly simulating it, which is a significant advantage, since a new library can be generated in a matter of hours whenever a new hadronic interaction model becomes available or a new experimental site needs to be evaluated. Also, all the terms in the sum can be made independent from each other allowing an efficient parallelization of the calculation.

\section{\label{sec:crepes_tasting}Testing CREPES}
\subsection{Vertical shower}

To test whether the CREPES algorithm performs as expected, we compare its predictions against a full ZHAireS simulation. We will compare its performance in the 30 to 350 MHz frequency band as all ground based radio experiments use a sub-range of it. We simulate a vertical 1 EeV proton shower developing in a standard atmosphere with a horizontal magnetic field oriented along the $x$ axis. This geometry represents the most demanding test case for the algorithm, since the observer is then located inside the shower itself, where the artifacts introduced by any approximation are most visible.

For the CREPES algorithm we need the distributions $w(\tau)$ and $\mathbf{u}(\tau)$, obtained directly from simulation, while $N(z)$ is simply given by a Gaisser-Hillas profile fitted to the electron and positron longitudinal development. To sample the distributions we selected 20 showers with longitudinal profiles similar to that of the shower whose radio emission we wished to evaluate. In each of the showers, every 100 g/cm$^2$ of slant depth, we saved the arrival time, direction, position, and weight of every electron and positron crossing a plane perpendicular to the shower axis. Averaging over 20 showers allows both the individual showers to be simulated more rapidly with thinning and yields a more stable shape for the resulting distributions, which are then interpolated between the sampled depths when evaluating the integral of eq.~\eqref{eq:crepes-algorithm}. Examples of these distributions are shown in Appendix~\ref{sf_evo}.

\begin{figure}[!ht]
\includegraphics{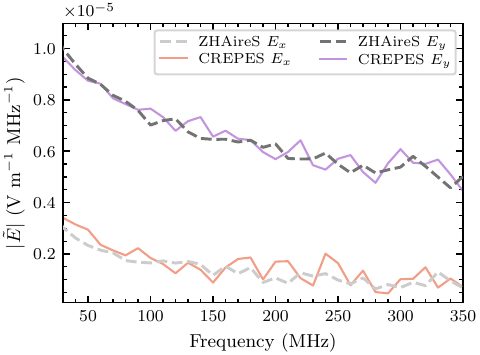}
\caption{\label{fig:vertical_spectrum_100} Comparison between ZHAireS and CREPES of the electric field spectrum from a 1 EeV proton vertical shower for an antenna located at (100,0) m at sea level altitude.
}
\end{figure}

Figures~\ref{fig:vertical_spectrum_50} and \ref{fig:vertical_spectrum_100} compare the ZHAireS simulation with the CREPES calculation for observers located at (50, 50, 0) m and (100, 0, 0) m, respectively. The agreement between the two calculations across the 30 to 350 MHz band is good, with no systematic trend across frequencies and no disagreement in the overall normalization of the pulse in either polarization. For both observers the peak electric field differs by less than 2.3\% between ZHAireS and CREPES. The performance of the algorithm degrades as the observer approaches the shower core: at 20 m from the core we find a 6.31\% difference in peak electric field, although the spectral shape continues to match well. This degradation is to be expected, since an observer close to the core lies within the shower itself, where the source points contributing to the pulse are close to the observer and the time differences between adjacent bins become considerable. Denser look-up tables can in principle be constructed to improve performance in this regime at the cost of sampling a larger number of showers in order to keep the number of particles per bin statistically relevant.

\begin{figure}[!ht]
\includegraphics{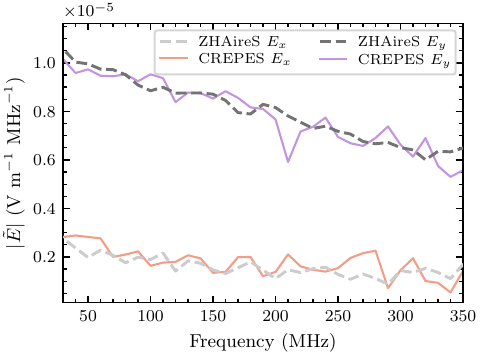}
\caption{\label{fig:vertical_spectrum_50} Comparison between ZHAireS and CREPES of the electric field spectrum from a 1 EeV proton vertical shower for an antenna located at (50,50) m at sea level altitude.
}
\end{figure}

\subsection{\label{sec:crepes_tasting80}Interpolating an 80\textdegree{} inclined shower}

We can now reproduce the radio emission from a shower with CREPES but this approach will still be slow if for every shower that we want to reconstruct we need to run several particle showers to then calculate the radio emission. The desired behavior would be to be able to simulate the edge cases of different geometries and interpolate between them to obtain the desired shower development. We will now show that this is indeed possible.

\begin{figure*}[!ht]
\includegraphics[width=\linewidth]{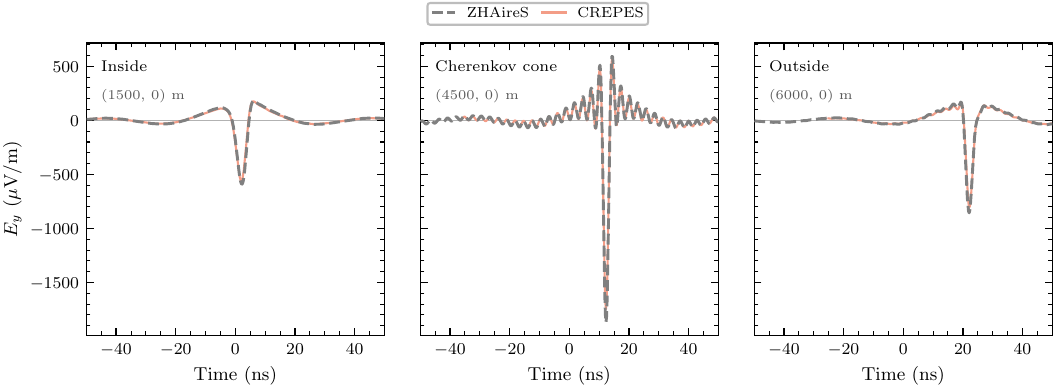}
\caption{\label{fig:80_degree_comparison} Y-polarization (geomagnetic) of the electric field of a 1 EeV proton shower at 80 degree zenith with a vertical magnetic field of 50 $\mu$T for ZHAireS and CREPES. Three cases are shown (from left to right): observers inside, at, and outside the Cherenkov cone. Only a global shift has been applied to all traces to align results.
}  
\end{figure*}

We now shift our focus to a more realistic scenario for the Pierre Auger Observatory where, with the Radio Detector (RD) recently deployed \cite{Horandel:2025xmh}, they are measuring radio signals from inclined showers \cite{Horandel:2025skz}. As the objective of this section is to verify the calculation we will still use the 30-350 MHz frequency band despite the RD measuring in the 30-80 MHz band. We run two edge cases for an 80\textdegree{} shower initiated by a 1 EeV proton: showers with $X_{max}$ at around 700 \grammage{} and 800 \grammage{}. Inclined showers develop in less dense atmosphere and particles drift further away in the same amount of grammage, so to mitigate this effect instead of sampling the particle distributions every 100 \grammage{} we do it every 25 \grammage{}.

The shower simulated with ZHAireS is a 1 EeV proton that has its shower maximum at 740 \grammage{} and no selection apart from the $X_{\textrm{max}}$ value was done. For the longitudinal profile we use the data available from the simulation. For $w(z',\tau')$ and $\textbf{u}(z',\tau')$ we interpolate linearly in depth between the distributions of the edge cases previously simulated. Fig.~\ref{fig:80_degree_comparison} shows the observed electric field for an observer inside, at and outside the Cherenkov cone filtered between 30 and 350 MHz only for the geomagnetic polarization as it is the dominant one in this geometry. The agreement between ZHAireS and CREPES is remarkable and their peak electric fields differs by 4.68\%, 6.58\% and 5.93\% for the observer inside, at and outside the Cherenkov cone, respectively. We note that there has only been a global shift between CREPES and ZHAireS, meaning that the arrival time from the detailed simulation and CREPES match perfectly.

\section{\label{sec:summary}Summary \& Outlook}

We have first reviewed what are the conditions that dictate the coherent radio emission of extensive air showers and, on this basis, we have visualized and studied the particle distributions of electrons and positrons at shower maximum. We have seen that these distributions are well characterized in a logarithmic scale in terms shower front delays, radial position and phi sector of the shower front. As shown in Figs.~\ref{fig:azimuthal_dist_at_xmax_vertical} and \ref{fig:azimuthal_dist_at_xmax_80}, the azimuthal dependence of these distributions cannot be neglected if the radio pulse is to be computed correctly. Evaluating the effective current terms (Figs.~\ref{fig:uy_velocity_distribution} and \ref{fig:ch_excess_distribution}) further shows that a one dimensional model would work for the geomagnetic emission while a three-dimensional model would be required for the Askaryan emission.

Based on the results from Section~\ref{sec:shower_anatomy} we have introduced the CREPES algorithm to calculate the radio emission from electron and positron distributions. The approach retains the azimuthal structure of these distributions and computes the total vector potential first, rather than the electric field, as done in the track-based calculation of ZHS~\cite{Alvarez-Muniz:2010wjm}, which is believed to give more stability to the calculation. For a vertical shower, relative peak field discrepancies are below 7\% and frequency spectra are are well described in the 30 to 350 MHz band, where ground based radio experiments operate. The same level of agreement is preserved for inclined showers, and it is maintained when the particle distributions are interpolated in order to obtain a shower geometry which was never sampled.

CREPES provides a framework in which only particle showers need to be sampled to obtain the radio emission efficiently, reducing the computation time of the radio emission from days to seconds once the particle distributions are obtained, with further speed-up expected from a GPU implementation. This opens the possibility of fitting shower profiles directly to data and of addressing optimization problems formulated in terms of the radio emission, such as the design of array layouts. A differentiable version of CREPES is in preparation, and further refinements in the sampling of the integral and of the quantities entering it are expected to improve the accuracy, although these were not the main focus of the present work.

\section*{Acknowledgments}
J. Ammerman-Yebra and H. Schoorlemmer are funded by the European Union. Views and opinions expressed are however those of the author(s) only and do not necessarily reflect those of the European Union or the European Research Council Executive Agency. Neither the European Union nor the granting authority can be held responsible for them. This work is supported by ERC grant (CR-INTERFEROMETRY 101170979). 


\appendix
\section{\label{sf_evo} Shower front evolution}

\begin{figure}[!ht]
\includegraphics[width=\linewidth]{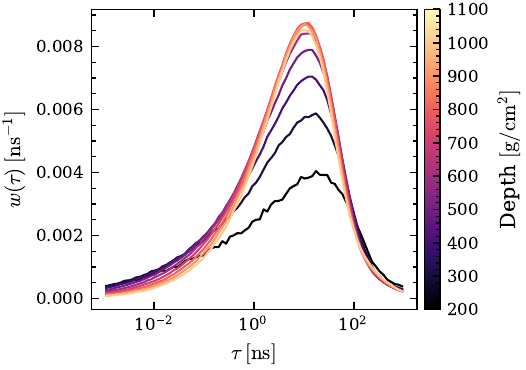}
\caption{\label{fig:distribution_sfd_depths} Evolution with depth of the shower front delay distribution for a vertical 1 EeV proton shower run with ZHAireS.
}  
\end{figure}

\begin{figure}[!ht]
\includegraphics[width=\linewidth]{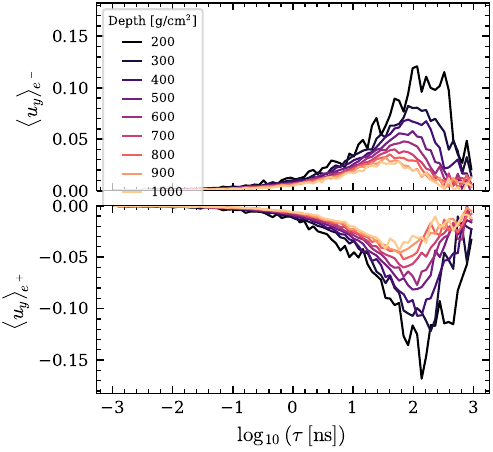}
\caption{\label{fig:uy_distribution_sfd} Evolution with depth of the drift velocity direction of electrons and positrons as a function of shower front delays. Values have been sampled from a vertical 1 EeV proton shower run with ZHAireS.
}  
\end{figure}

For a one-dimensional shower, in which all particles are projected onto the shower
axis (the case of eq.~\eqref{eq:1d-frape}), the distributions involved can be easily
visualized. Fig.~\ref{fig:distribution_sfd_depths} shows the evolution of the shower
front delay distribution with depth for a 1~EeV vertical proton shower. Fig.~\ref{fig:uy_distribution_sfd} shows the evolution of the distribution of the
drift component of the velocity direction (for this geometry $u_y$), given separately for electrons and
positrons, which deflect in opposite directions due to their opposite charges.

\vspace{4cm}

\bibliography{main}

\end{document}